\documentstyle [aps,pre,epsf] {revtex}

\newcommand{\cH} {{\cal H}}
\newcommand{\cO} {{\cal O}}
\newcommand{\eg} {{\it e.g., }}
\newcommand{\eps} {\epsilon}
\newcommand{\half} {\frac{1}{2}}
\newcommand{\ie} {{\it i.e., }}
\newcommand{\lp} {l_{\rm p}}
\newcommand{\ph} {\varphi}
\newcommand{\rmd} {{\rm d}}
\newcommand{\rme} {{\rm e}}

\newcommand{\vect} {{\bf t}}
\newcommand{\vecu} {{\bf u}}

\def\tr{\mathop{\mbox{Tr}}}

\begin{document}

\draft

%------------------------------------------------------
% 1st page
%------------------------------------------------------

\title {Binding of molecules to DNA and other semiflexible polymers}

\author {Haim Diamant\footnote{Present address: The James Franck
Institute, The University of Chicago, 5640 South Ellis Avenue,
Chicago, IL 60637.}
   and David Andelman}
\address {School of Physics and Astronomy,
         Raymond and Beverly Sackler Faculty of Exact Sciences,\\
         Tel Aviv University, Ramat Aviv, 69978 Tel Aviv, Israel}

\date{\today}

\maketitle

\begin{abstract}

A theory is presented for the binding of small molecules such as
surfactants to semiflexible polymers. The persistence length is
assumed to be large compared to the monomer size but much smaller
than the total chain length. Such polymers (\eg DNA) represent an
intermediate case between flexible polymers and stiff, rod-like
ones, whose association with small molecules was previously
studied. The chains are not flexible enough to actively
participate in the self-assembly, yet their fluctuations induce
long-range attractive interactions between bound molecules. In
cases where the binding significantly affects the local chain
stiffness, those interactions lead to a very sharp, cooperative
association. This scenario is of relevance to the association of
DNA with surfactants and compact proteins such as RecA. External
tension exerted on the chain is found to significantly modify the
binding by suppressing the fluctuation-induced interaction.

\end{abstract}


%\vspace{3.5cm}\noindent
%Submitted to

%\setlength {\baselineskip} {20pt}
%\pagebreak
%--------------------------------------------------------------

\section{Introduction}
%---------------------
\label{introduction}

Aqueous solutions containing polymers and small associating
molecules such as folded proteins and amphiphiles (surfactants)
are commonly found in biological systems and industrial
applications. As a result, extensive efforts have been devoted in
the past few decades to the study of polymer--surfactant
interactions \cite{ps_book1,ps_book2}. In addition, there has been
growing interest in the interactions between DNA macromolecules
and surfactants, lipids and short polyamines \cite
{Hayakawa,Shirahama,Sergeyev1,Sergeyev5,Khokhlov,Sergeyev6,Reimer,Bhatta}.
These interactions are relevant to various biochemical
applications such as DNA extraction and purification
\cite{Sergeyev6,Reimer,Bhatta} and genetic delivery systems
\cite{delivery}. Association of folded proteins (\eg RecA) with
DNA plays a key role in genetic regulatory mechanisms. Structural
details of this association have been studied in recent
experiments \cite{Chatenay,Feingold}.

Recently, we have presented a general theory for the self-assembly
in aqueous solutions of polymers and smaller associating molecules
\cite{ourEPL,ourMM}. Two different scenarios emerge, depending on
the flexibility of the polymer. If the polymer is flexible enough,
it actively participates in the self-assembly, resulting in mixed
aggregates jointly formed by the two species. The polymer
conformation changes considerably upon self-assembly but remains
extended on a global scale, as the chain undergoes only {\em
partial collapse} \cite{ourEPL,ourMM,deGennes_partial}. On the
other hand, if the polymer is stiff, partial collapse is
inhibited.
%and the polymer may either have the same statistics
%upon self-assembly, or undergo {\em complete coil-to-globule
%collapse}.

The criterion determining the `flexible' {\it vs.} `stiff'
scenarios concerns the polymer statistics on a mesoscopic length
scale characterizing correlations in the solution (usually a few
nanometers). It was found \cite{ourEPL,ourMM}
that the flexible (stiff) scenario holds
if the exponent $\nu$, relating the number of monomers $N$ to the
spatial size $R$ they occupy, $R\sim N^\nu$, is smaller (larger)
than $2/d$ on that length scale ($d$ being the dimensionality).
This distinction is analogous to the one made in the critical
behavior of certain disordered systems \cite{Fisher,Harris}
--- if the critical exponent $\nu$ of a system satisfies
$\nu<2/d$, the critical behavior would be smeared by impurities
(in analogy to the partial collapse), whereas if $\nu>2/d$, the
critical point remains intact. Indeed, neutral flexible polymers
in three dimensions, having $\nu\simeq 3/5<2/3$, are found by
scattering experiments to associate with surfactants in the form
of a `chain of wrapped aggregates' \cite{Cabane,Reed}. On the
other hand, stiff DNA molecules, having $\nu=1$ on the relevant
length scale, are found to either remain unperturbed by surfactant
binding \cite{Sergeyev5,Reimer}, or undergo a discontinuous
coil-to-globule transition \cite{Sergeyev1}, provided the chain is
much longer than the persistence length.

In previous publications \cite{ourEPL,ourMM} we concentrated on
the flexible case and the corresponding partial collapse, where
the polymer degrees of freedom play an important role. In the
opposite extreme limit of stiff, rod-like molecules, the
conformational degrees of freedom of the polymer can be neglected
and the chain may be regarded as a linear `binding substrate'.
Models for stiff polymers, inspired by the Zimm-Bragg theory
\cite{ZimmBragg}, treat the bound molecules as a one-dimensional
lattice-gas (or Ising) system with nearest-neighbor interactions
\cite{Ising_models}. They have been widely used to fit
experimental binding isotherms for polyelectrolytes and oppositely
charged surfactants \cite{review_ps_charge}. Recently, more
detailed electrostatic models have been proposed for the
interaction between rod-like polyelectrolytes and oppositely
charged surfactants \cite{Levin,Colby}.
In addition, a theoretical work focusing on the {\it specific} 
binding of proteins to 
DNA has been recently presented \cite{Rudnick},
treating a pair of bound proteins as geometrically constraining 
inclusions on the DNA chain.

In the current work we address the intermediate case of {\em
semiflexible} polymers. The polymer we consider is stiff in the
sense defined above, \ie its persistence length, $\lp$, exceeds
several nanometers and, hence, the polymer is characterized by
$\nu=1>2/3$ on that length scale. The total chain length, however,
is considered to be much larger than $\lp$, and the entire polymer
cannot be regarded, therefore, as a single rigid rod. This case
corresponds, in particular, to experiments on long DNA molecules
\cite{Hayakawa,Shirahama,Sergeyev1,Sergeyev5,Khokhlov,Sergeyev6,Reimer,Bhatta},
whose persistence length is typically very large (of order 50 nm),
but much smaller than the total chain length (which is usually
larger than a micron) \cite{Gorelov}. We argue that such an
intermediate system may, in certain cases, be governed by
different physics. Although the polymer is too stiff to change
conformation and actively participate in the self-assembly, its
degrees of freedom induce attractive correlations between bound
molecules. Those fluctuation-induced correlations are weak but
have a long spatial range (of order $\lp$) and, hence, may
strongly affect the binding thermodynamics.

%The problem addressed in the current work is different in three 
%important aspects.
%(i) The bound molecules are not assumed to be sequence-specific, static
%inclusions, but are described by thermodynamic degrees of freedom.
%Our theory, therefore, is mainly relevant to {\it non-specific}
%bound agents such as surfactants and RecA proteins.
%(ii) The mechanism proposed in Ref.~\cite{Rudnick} relies on
%external tension applied to the DNA chain. 
%The effects resulting from our
%analysis are sensitive, indeed, to external tension, but are 
%present also in the absence of tension.
%(iii) The calculations of Ref.~\cite{Rudnick} are restricted 
%to pair interactions. It is shown in subsequent sections, however,
%that the fluctuation-induced interaction between bound molecules
%has significant many-body contributions as well.

The model is presented in Sec.~\ref{model}. Bound molecules are
assumed to modify the local features of polymer conformation, \eg
change its local stiffness. In the limit of weak coupling, our
model reduces to the Kac-Baker model \cite{Lieb,Kac,Baker}, which
is solvable exactly. This limit is discussed in Sec.~\ref{weak}.
Although turning out to be of limited interest in practice, the
weak-coupling limit provides insight into the mechanism of
association, and helps us justify further approximations. Section
\ref{strong} presents a mean-field calculation for an arbitrary
strength of coupling. This analysis leads to our main conclusions,
and in Sec.~\ref{sec_tension} it is extended to polymers under
external tension. The results are summarized in
Sec.~\ref{conclusions}, where we also discuss several relevant
experiments involving DNA and point at future directions.

\section{The Model}
%------------------
\label{model}

Small molecules bound to stiff polymers are commonly modeled as a
one-dimensional lattice gas (or Ising system) \cite{Ising_models}.
Each monomer serves as a binding site, which can either
accommodate a small molecule or be empty, and the surrounding
dilute solution is considered merely as a bulk reservoir of small
molecules. In the current work we stay at the level of a
one-dimensional model, assuming that the polymer is still quite
(yet not infinitely) stiff, \ie the persistence length is much
larger than the monomer size. In addition, a dilute polymer limit
is assumed, where inter-chain effects can be neglected. We focus
on the effect of introducing the polymer degrees of freedom and,
hence, seek a simple meaningful coupling between the polymer and
the bound `lattice gas'.

A polymer configuration is defined by a set of vectors,
$\{\vecu_n\}_{n=1\ldots N}$, specifying the lengths and
orientations of the $N$ monomers. In addition, each monomer serves
as a binding site which can be either empty ($\ph_n=0$) or
occupied by a small molecule ($\ph_n=1$). A configuration of the
entire system is defined, therefore, by specifying
$\{\vecu_n,\ph_n\}_{n=1\ldots N}$.

Since the polymer is assumed to be locally stiff, a natural choice
would be to couple $\ph_n$ with the square of the local chain
curvature, $\ph_n(\vecu_{n+1}-\vecu_n)^2$, thus modifying the
local chain stiffness. However, in the usual Kratky-Porod
worm-like-chain model of semiflexible polymers \cite{WLC},
chain segments are taken as rigid rods of fixed length
($|\vecu_n|=\mbox{const}$), and each squared-curvature term
contains only one degree of freedom (\eg the angle $\theta_n$
between $\vecu_n$ and $\vecu_{n+1}$). Consequently, this coupling,
$\ph_n\cos\theta_n$, would leave $\{\ph_n\}$ uncorrelated, leading
merely to a trivial shift in the chemical potential of bound
molecules \cite{Rudnick2}.
One option to proceed is to consider higher-order
extensions of the worm-like-chain Hamiltonian, involving three consecutive
monomers. This will introduce correlations between bound molecules
at different sites. 

We take, however, a simpler route and modify
the worm-like-chain model by allowing the monomer length to fluctuate.
This modification was originally presented by Harris and Hearst
\cite{HH}, using a single global constraint for the average
chain length. The modified model was shown to successfully reproduce
the results of the Kratky-Porod model as far as thermodynamic averages
(\eg correlation functions, radius of gyration) were concerned.
It was less successful, however, in recovering more detailed statistics 
of the worm-like chain (\eg distribution function, form factor),
particularly in the limit of large stiffness.
The Harris-Hearst model was later refined by Lagowski {\it et al.} 
\cite{Lagowski}
and Ha and Thirumalai \cite{HT,HT_tension}, replacing the single global
constraint by a set of local constraints for the average segment lengths.
This further modification was shown to be equivalent to a 
stationary-phase approximation for the chain partition function, 
yielding reliable results for average quantities, as well as more
detailed statistics \cite{HT}.
We note that a similar
approach was used in a recent model of semiflexible polymer
collapse \cite{Parsegian}.
It should be borne in mind that, despite its success in the past,
the constraint relaxation remains essentially an uncontrolled
approximation. In the current work we restrict ourselves to
thermodynamic averages, such as monomer-monomer correlations and
free energies, for which the modified model with a single
global contraint can be trusted.

Thus, the rigid constraints of the original Kratky-Porod model, 
$u_n^2=1$, are relaxed into thermodynamic-average ones, 
$\langle u_n^2\rangle=1$, where the
mean-square monomer size is taken hereafter as the unit length.
Using the modified model for the chain,
each $\ph_n(\vecu_{n+1}-\vecu_n)^2$ term involves two consecutive
monomers (and not merely the angle between them), leading to a
meaningful coupling between binding and polymer conformation.

The partition function of the combined system of polymer and bound
molecules is written, therefore, as
\begin{eqnarray}
  Z &=& \tr_{\{\ph_n=0,1\}} \int\prod_{n=1}^N \rmd
  \vecu_n \exp(-\cH)
  \nonumber\\
  \cH &=& \frac{3}{4}\lp \sum_{n=1}^{N-1}
  (1+\eps\ph_n) (\vecu_{n+1}-\vecu_n)^2 + \sum_{n=1}^N
  \lambda_n u_n^2 - \mu\sum_{n=1}^N\ph_n.
\label{Z1}
\end{eqnarray}
In Eq.~(\ref{Z1}) $\lp$ is the persistence length of the bare
chain, characterizing its intrinsic stiffness. It is assumed to be
much larger than the monomer size, $\lp\gg 1$. The coupling is
introduced through the stiffness term, assuming that a bound
molecule modifies the local stiffness by a fraction $\eps>-1$,
which may be either negative or positive but cannot change the
positive sign of the overall stiffness term
\cite{negative_stiffness}. The second term contains a set of
multipliers, $\lambda_n$, to be chosen so that the constraints
$\langle u_n^2\rangle=1$ are satisfied. However, replacement of
the entire set $\{\lambda_n\}$ by a single multiplier $\lambda$
can be shown to yield a non-extensive correction \cite{HT}, which
becomes negligible in the limit $N\rightarrow\infty$. Hence, we
use hereafter a single multiplier, $\lambda$. Finally, the system
is assumed to be in contact with a reservoir of solute molecules.
The last term in Eq.~(\ref{Z1}) accounts for this contact along
with any other factors which couple linearly to the degree of
binding. Typically, $\mu$ contains the chemical potential of the
solute reservoir and the direct energy of solute molecule--monomer
binding. (All energies in this work are expressed in units of the
thermal energy, $k_{\rm B}T$.) Note that we have not included in
Eq.~(\ref{Z1}) any direct short-range (\eg nearest-neighbor)
interactions between bound molecules. Thus, all interactions in
the model arise from the coupling to the polymer degrees of
freedom. Short-range interactions between bound molecules do exist
in physical systems. Yet, in the limit of $\lp\gg 1$ and 
$|\eps|\gtrsim 1$, which is of interest to the current work, 
such direct interactions have a minor effect on binding, as is
demonstrated in the following sections. Hence, we omit them for
the sake of brevity.

As a reference, let us start with the previously studied
partition function of the bare polymer \cite{HT},
\begin{equation}
  Z_{\rm p} = \int\prod_n\rmd\vecu_n \exp[ -\frac{3}{4}\lp \sum_n
  (\vecu_{n+1}-\vecu_n)^2 - \lambda\sum_n u_n^2].
\label{Zp1}
\end{equation}
It is a Gaussian integral which can be calculated either by
transforming it to Fourier space and integrating, or by analogy to
the path integral of a three-dimensional quantum oscillator
\cite{QM}. The result in the limit $N\rightarrow\infty$ and for
$\lp\gg 1$ is
\begin{equation}
  Z_{\rm p}^{1/N} = \left(\frac{4}{3\pi\lp}\right)^{3/2}
  \exp\left(3-\sqrt{3\lambda/\lp}\right).
\label{Zp2}
\end{equation}
The multiplier $\lambda$ can now be determined according to
\begin{equation}
  -\frac{1}{N} \frac{\partial\log Z_{\rm p}}{\partial\lambda} =
  \langle u_n^2\rangle_{\rm p} = 1 \ \ \Longrightarrow \ \ \lambda =
  \frac{3}{4\lp},
\label{lambda_sol}
\end{equation}
where $\langle\cdots\rangle_{\rm p}$ denotes a thermal average
over the bare chain statistics (\ie using $Z_{\rm p}$). The
corresponding free energy per monomer (in the ensemble of {\em
constrained} $\vecu_n$) is
\begin{equation}
  f_{\rm p} = -\frac{1}{N}\log Z_{\rm p} - \lambda
  = \frac{3}{2}\log\lp  + \frac{3}{4\lp} + \mbox{const}.
\label{fp}
\end{equation}

Various correlations in the bare chain can be calculated. The pair
correlation between segment vectors along the chain sequence is
\begin{equation}
  \langle\vecu_m\cdot\vecu_n\rangle_{\rm p} = \rme^{-|m-n|/\lp},
\label{segment_correlation}
\end{equation}
which explains why the parameter $\lp$ has been defined as the
persistence length. Two higher-order pair correlations are
calculated as well:
\begin{eqnarray}
  g_1 &\equiv& \langle(\vecu_{n+1}-\vecu_n)^2\rangle_{\rm p}
  = \frac{2}{\lp} + \cO(\lp^{-2})
  \nonumber\\
  g_2(m,n) &\equiv& \langle (\vecu_{m+1}-\vecu_m)^2(\vecu_{n+1}-\vecu_n)^2
  \rangle_{\rm p} - g_1^2
  = \frac{8}{3\lp^3}\rme^{-2|m-n|/\lp} + \cO(\lp^{-4}),
\label{g12}
\end{eqnarray}
and will be of use in the next section, where we re-examine the
coupled system.

\section{Weak Coupling}
%----------------------
\label{weak}

Let us return to the full partition function (\ref{Z1}),
which can be equivalently written as
\begin{equation}
  Z = Z_{\rm p} \tr_{\{\ph_n\}} \exp(\mu\sum_n\ph_n)
  \left\langle \exp[ -\frac{3\lp\eps}{4} \sum_n
  \ph_n(\vecu_{n+1}-\vecu_n)^2 ] \right\rangle_{\rm p}.
\label{Z2}
\end{equation}
First we consider the weak-coupling limit, $|\eps|\ll 1$, where
the partition function (\ref{Z2}) can be treated by a cumulant
expansion. In this limit the model becomes analogous to the
exactly solvable Kac-Baker model \cite{Lieb,Kac,Baker}, 
and we show that identical
results are derived from a simple mean-field calculation. We then
use this observation to justify a mean-field calculation for an
arbitrary value of $\eps$.

A cumulant expansion of Eq.~(\ref{Z2}) to 2nd order in $\eps$
leads to
\begin{equation}
  Z \simeq Z_{\rm p} \tr_{\{\ph_n\}} \exp \left[ \left( \mu -
  \frac{3\lp\eps}{4}g_1 \right) \sum_n\ph_n + \half\left(
  \frac{3\lp\eps}{4}\right)^2 \sum_{m,n} g_2(m,n)\ph_m\ph_n \right],
\label{Z3}
\end{equation}
where the correlations $g_1$ and $g_2$ were defined in
Eq.~(\ref{g12}). Substituting expressions (\ref{g12}), the
partition function is decoupled into a polymer contribution and an
effective contribution from the bound solute molecules,
\begin{eqnarray}
  Z &\simeq& Z_{\rm p} Z_{\rm s} = Z_{\rm p} \tr_{\{\ph_n\}}
  \exp(-\cH_{\rm  s})
  \nonumber\\
  \cH_{\rm s} &=& \half\sum_{m\neq n} V_{mn}\ph_m\ph_n
  - \hat{\mu} \sum_n\ph_n,
\label{Z4}
\end{eqnarray}
where
\begin{eqnarray}
  V_{mn} &\equiv& -\frac{3\eps^2}{2\lp} \rme^{-2|m-n|/\lp}
  \nonumber\\
  \hat{\mu} &\equiv& \mu - \frac{3\eps}{2} + \frac{3\eps^2}{4\lp}.
\label{vmn}
\end{eqnarray}

The introduction of the polymer degrees of freedom and their
coupling to the binding ones have led to two effects, as compared
to previous lattice-gas theories. First, there is a shift in the
chemical potential, $\mu\rightarrow\hat{\mu}$. This is equivalent
to an effective change in the affinity between the small molecules
and the chain. As expected, if binding strengthens the local
stiffness of the chain ($\eps>0$), the affinity is reduced (\ie
the isotherm is shifted to higher chemical potentials), whereas if
it weakens the stiffness ($\eps<0$), the shift is to lower $\mu$.
The second, more interesting effect is that bound molecules
experience an attractive potential, $V_{mn}$, along the chain. The
amplitude of this effective interaction is small
($\sim\eps^2/\lp$), but its range is large
--- of order $\lp$. When $\lp$ is increased there are two opposing
consequences
--- the interaction amplitude diminishes, while the interaction
range is extended. The overall effect on the thermodynamics of
binding, therefore, has to be checked in detail.

\subsection{Analogy with the Kac-Baker Model}
%--------------------------------------------

The effective Hamiltonian of the bound solute, $\cH_{\rm s}$, is a
lattice-gas version of the Kac-Baker model \cite{Lieb,Kac,Baker},
which is exactly solvable. Moreover, the procedure relevant to our
semiflexible polymer, \ie increasing $\lp$ while keeping
$1\ll\lp\ll N$, is precisely the one studied in detail by Kac and
Baker. Their results, as applied to our binding problem, can be
summarized as follows. For any finite $\lp$, the bound molecules
are always in a disordered state along the polymer chain, as in
any one-dimensional system with finite-range interactions.
Consequently, the binding isotherm, \ie the binding degree
$\ph\equiv\langle\ph_n\rangle$ as function of $\mu$ (see, \eg
Fig.~\ref{fig_mf}a), is a continuous curve. However, in the limit
$\lp\rightarrow\infty$, taken {\em after} the infinite-chain limit
$N\rightarrow\infty$, there is a critical value of coupling above
which the binding exhibits a discontinuous (1st-order) transition.
According to Baker's rigorous calculation \cite{Baker}, the
critical value of the potential amplitude multiplied by $\lp$
(equal, in our case, to $3\eps_{\rm c}^2/2$) is 4, \ie
\begin{equation}
  \eps_{\rm c}^\pm = \pm\sqrt{8/3} \simeq \pm 1.63.
\label{eps_c1}
\end{equation}
Note that the symmetry with respect to the sign of $\eps$ is
merely an artificial consequence of our 2nd-order expansion,
Eq.~(\ref{Z3}). In general, the results should not be the same if
the stiffness is weakened ($\eps<0$) or strengthened ($\eps>0$),
as is demonstrated in Sec.~\ref{strong}.

The negative critical value in Eq.~(\ref{eps_c1}), $\eps_{\rm
c}^-\simeq -1.63$, lies outside the range of validity of the
original polymer binding model, $\eps>-1$ [cf. Eq.~(\ref{Z1})].
The positive value, $\eps_{\rm c}^+\simeq 1.63$, does not satisfy
the assumption of weak coupling, $|\eps|\ll 1$, which have led to
the analogy with the Kac-Baker model in the first place. Thus, the
sharp binding isotherms obtained from the Kac-Baker model for
$|\eps|>\eps_{\rm c}$ do not apply, strictly speaking, for our
polymer binding problem. The weak-coupling calculation does
demonstrate, however, how fluctuations in polymer conformation
induce long-range attraction between bound molecules. This basic
feature is expected to remain when one considers stronger
coupling, $|\eps|>1$, and the resulting many-body terms omitted in
Eq.~(\ref{Z3}). This is further discussed in the following
sections.

Finally, the polymers we consider have a large but finite $\lp$.
For example, the persistence length of a DNA macromolecule is
typically of order 50--100 nm, whereas the length of a single base
pair is $0.34$ nm. Hence, $\lp$ is of order $10^2$ (in units of
monomer length) . It is worth checking to what extent the
sharpness of binding in the Kac-Baker model for $|\eps|>\eps_{\rm
c}$ is affected by finite $\lp$. For this purpose, let us define a
{\it cooperativity parameter} for the binding, measuring the
maximum slope of the binding isotherm,
\begin{equation}
  C \equiv \left. \frac{\partial\ph}{\partial\mu}
  \right|_{\rm max} - \frac{1}{4}.
\label{cooperativity1}
\end{equation}
This parameter is equivalent to the zero magnetic field
susceptibility in the analogous spin system, and is commonly
measured from the slope of binding isotherms obtained in
potentiometric experiments \cite{ps_book1,ps_book2}. It has been
defined in Eq.~(\ref{cooperativity1}) so as to yield zero for
vanishing interaction ($\eps=0$) and diverge at a critical point.
(In the current weak-coupling limit, the maximum slope is obtained
for $\langle\ph\rangle=1/2$.) Given $\lp$ and $\eps$, the
cooperativity is numerically calculated using Kac's exact solution
\cite{Lieb,Kac}, as is explained in the Appendix.
Figure~\ref{fig_Kac} presents the results for $\lp=10$ and 50. For
$\lp=50$ the binding becomes highly cooperative for
$|\eps|>\eps_{\rm c}$. For even larger values of $\lp\sim 10^2$
(relevant, \eg to DNA) the binding will be hardly distinguishable
from that of an infinite $\lp$.

\subsection{Mean-Field Calculation}
%----------------------------------

In fact, the results of the Kac-Baker model in the limit
$N\rightarrow\infty, \lp\rightarrow\infty$, while keeping $\lp<N$,
can be also obtained from a simple mean-field calculation
\cite{Lieb,Schwartz}. The heuristic argument for this agreement is
the following: as $\lp$ is increased, the range of interaction is
extended and each bound molecule interacts with an increasing
number of neighbors. As a result, the averaging assumption
underlying the mean-field approximation is justified, and becomes
{\em exact} when the range of interaction is taken to infinity.
The correspondence between infinite-range models and mean field
was rigorously proved by Lebowitz and Penrose for a more general
class of potentials \cite{Lebowitz}.

Indeed, employing a mean-field approximation for the potential
(\ref{vmn}) in the limit of very large $\lp$,
\[
  \sum_{mn} V_{mn}\ph_m\ph_n \rightarrow -\frac{3\eps^2}{2\lp}
  \left(\sum_{mn}\rme^{-2|m-n|/\lp}\right) \ph^2
  \simeq -\frac{3\eps^2}{2}N\ph^2,
\]
where $\ph$ is an average, uniform binding degree,
we are led to the following mean-field free energy
per monomer:
\begin{equation}
  f = f_{\rm p} + f_{\rm s} \simeq
  f_{\rm p} + \ph\log\ph + (1-\ph)\log(1-\ph) -
  \frac{3\eps^2}{4}\ph^2 - \hat{\mu}\ph,
  \ \ \ \  \mbox{for~~} \lp\rightarrow\infty.
\label{f_mf1}
\end{equation}
It is easily verified that the critical point of this free energy
is $\eps_{\rm c}^2=8/3$, in agreement with the rigorous result,
Eq.~(\ref{eps_c1}). The cooperativity parameter can be calculated
as well from Eq.~(\ref{f_mf1}), yielding
\begin{equation}
  C = \frac{\eps^2}{4(\eps_{\rm c}^2-\eps^2)}, \ \ \ \ 
  \mbox{for~~} \lp\rightarrow\infty.
\label{C_mf1}
\end{equation}
This expression shows the usual critical behavior obtained from
mean-field theories, $C\sim|\eps-\eps_{\rm c}|^{-\gamma}$ with
$\gamma=1$. The dependence of $C$ on $\eps$ according to
Eq.~(\ref{C_mf1}) is shown by the solid line in
Fig.~\ref{fig_Kac}. The curves obtained from Kac's solution
approach it, as expected, when $\lp$ is increased. Recall
that expressions (\ref{f_mf1}) and (\ref{C_mf1}) correspond to the
original problem of bound molecules only in the limit of small
$\eps$.

\section{Strong Coupling}
%------------------------
\label{strong}

The interesting part of our theory requires $|\eps|\gtrsim 1$ and
thus limits the interest in the analogy to the Kac-Baker model.
Nevertheless, based on the heuristic argument given above, it is
reasonable to assume that, in the limit $\lp\gg 1$, the mean-field
approximation is good for larger values of $|\eps|$ as well
\cite{proof}. The preceding section, discussing the Kac-Baker
model in the weak-coupling limit, may be regarded, therefore, as a
justification for using the mean-field approximation for
one-dimensional models with large $\lp$ and $|\eps|\gtrsim 1$.
Applying a mean-field approximation to the binding degrees of
freedom $\ph_n$ in our starting point, Eq.~(\ref{Z1}), the
tracing over $u_n$ can be done exactly. The resulting free energy
is composed of the polymer free energy, $f_{\rm p}$, evaluated
with an effective persistence length,
$\lp\rightarrow\lp(1+\eps\ph)$, and the entropy of mixing for
$\ph$,
\begin{equation}
  f = \left.f_{\rm p}\right|_{\lp\rightarrow\lp(1+\eps\ph)}
  + \ph\log\ph + (1-\ph)\log(1-\ph) - \mu\ph.
\end{equation}
Using Eq.~(\ref{fp}), we obtain
\begin{equation}
  f = \ph\log\ph + (1-\ph)\log(1-\ph) +
  \frac{3}{2}\log[\lp(1+\eps\ph)] + \frac{3}{4\lp(1+\eps\ph)}
  - \mu\ph.
\label{f_mf2}
\end{equation}
For $\eps\ll 1$ and $\lp\gg 1$ this expression reduces, as
expected, to our previous result for the weak-coupling limit,
Eq.~(\ref{f_mf1}).

\begin{figure}
\centerline{ \epsfxsize=8.6cm \hbox{\epsffile{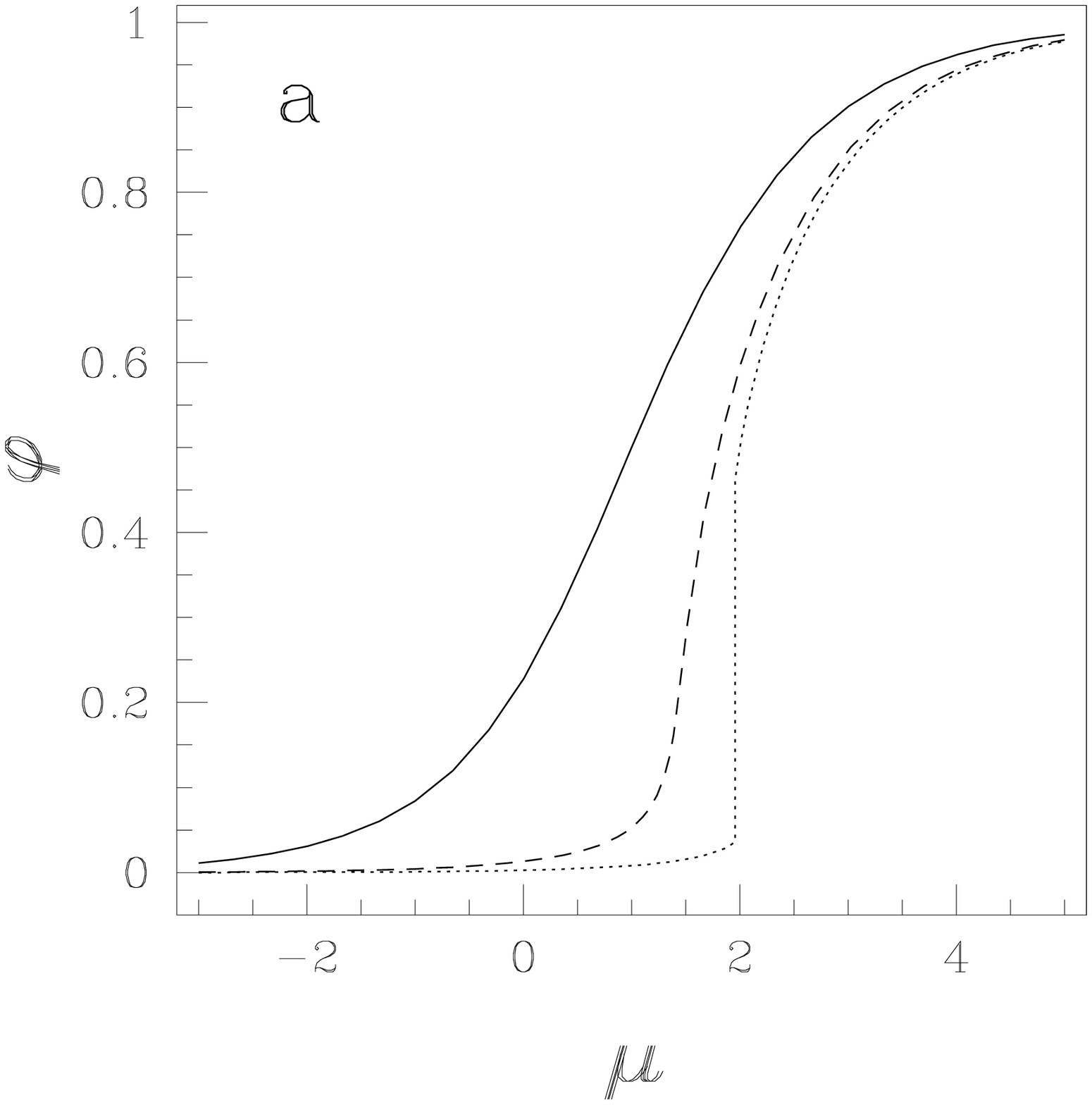}}
\epsfxsize=8.6cm \hbox{\epsffile{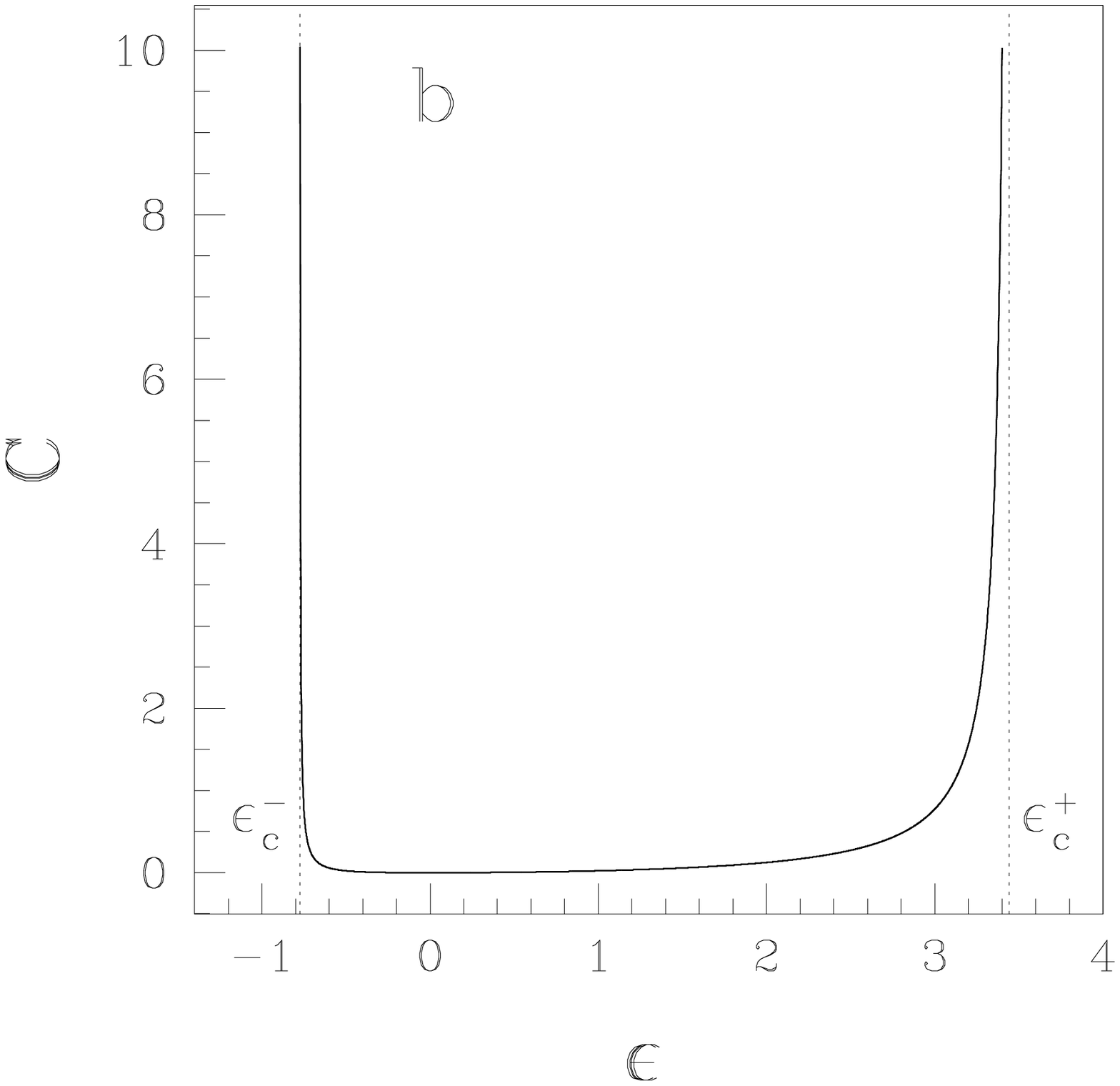}}}
%\centerline{
%\resizebox{0.6\textwidth}{!}{\includegraphics{fig2.ps}}}
\caption[]{(a) Binding isotherms obtained from the mean-field
theory [Eq.~(\ref{iso_mf2})] for three different values of $\eps$:
$\eps=1$ (solid line), $\eps=3$ (dashed) and $\eps=4$ (dotted),
the latter being beyond the critical point
$\eps_{\rm c}^+\simeq 3.44$. The chemical potential $\mu$ is
given in units of $k_{\rm B}T$.
(b) Binding cooperativity as function of $\eps$ according
to the mean-field calculation [Eq.~(\ref{C_mf2})]. The
cooperativity diverges at the two critical points $\eps_{\rm
c}^\pm=2(2\pm\sqrt{10})/3$ (dotted lines), beyond which binding
isotherms
exhibit a 1st-order transition [see dotted curve in (a)].}
\label{fig_mf}
\end{figure}

In the limit $\lp\gg 1$ the critical points of the free energy
(\ref{f_mf2}) are
\begin{equation}
  \eps_{\rm c}^- = \frac{2}{3} \left(2-\sqrt{10}\right)
  \simeq -0.775,\ \ \
  \eps_{\rm c}^+ = \frac{2}{3} \left(2+\sqrt{10}\right)
  \simeq 3.44,
\label{eps_c2}
\end{equation}
both of which lie within our general range of validity, $\eps>-1$.
(Note the loss of symmetry with respect to the sign of $\eps$,
which was a consequence of the weak-coupling approximation in
Sec.~\ref{weak}.) The corresponding critical chemical potentials
are
\begin{equation}
  \mu_{\rm c}^\pm = \frac{3\eps_{\rm c}^\pm(\eps_{\rm c}^\pm+2)}
  {4(\eps_{\rm c}^\pm+1)} - \log(\eps_{\rm c}^\pm + 1)
  \simeq \pm 1.67.
\label{mu_c}
\end{equation}
The binding isotherm, $\ph=\ph(\mu)$, as derived from
Eq.~(\ref{f_mf2}), satisfies
\begin{equation}
  \mu = \log\frac{\ph}{1-\ph} + \frac{3\eps}{2(1+\eps\ph)},
  \ \ \ \ \ \lp\gg 1.
\label{iso_mf2}
\end{equation}
Figure \ref{fig_mf}a shows three binding isotherms for three
different values of $\eps$ below and above the critical point.
The corresponding binding cooperativity is
\begin{equation}
  C = \frac {8(1+\eps)^2} {3(2+\eps)^2 (\eps-\eps_{\rm c}^-)
  (\eps_{\rm c}^+-\eps)} - \frac{1}{4}, \ \ \ \ \
  \lp\gg 1.
\label{C_mf2}
\end{equation}
As in Eq.~(\ref{C_mf1}), this expression exhibits the usual
mean-field critical behavior, $C\sim|\eps-\eps_{\rm c}|^{-\gamma}$
with $\gamma=1$. The dependence of $C$ on $\eps$ is plotted in
Fig.~\ref{fig_mf}b.

\begin{figure}
\centerline{ \epsfxsize=8.6cm \hbox{\epsffile{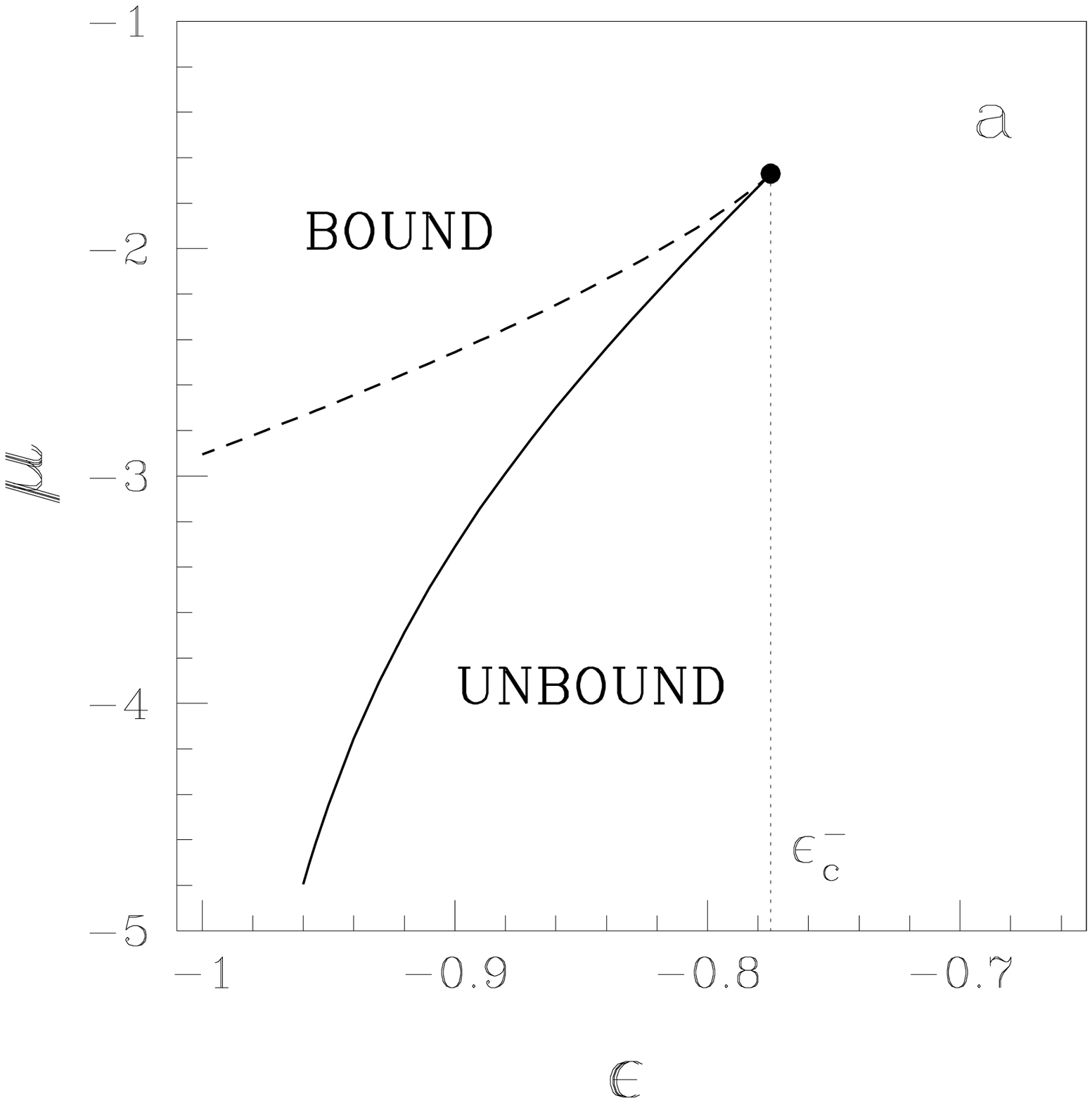}}
\epsfxsize=8.6cm \hbox{\epsffile{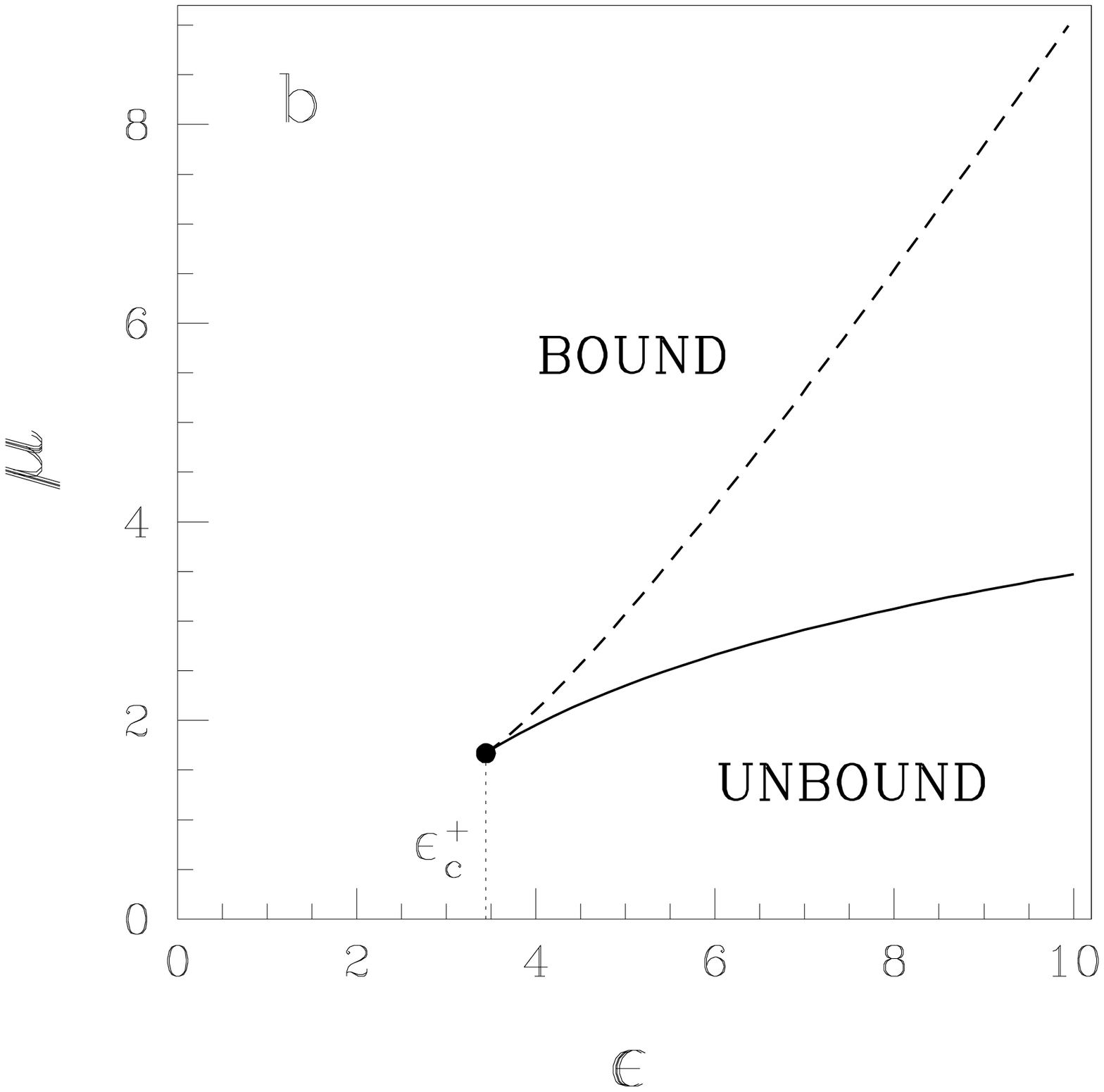}}}
%\centerline{
%\resizebox{0.5\textwidth}{!}{\includegraphics{fig3a.ps}}
%\resizebox{0.5\textwidth}{!}{\includegraphics{fig3b.ps}}}
\caption[]{Binding phase diagrams calculated from the free energy
(\ref{f_mf2}). (a) Phase diagram for stiffness-weakening binding
($-1<\eps<0$); (b) phase diagram for stiffness-strengthening
binding ($\eps>0$). Solid and dashed lines indicate the binodal
and spinodal, respectively. The lines meet at the critical points
$(\eps_{\rm c}^\pm,\mu_{\rm c}^\pm)$ given by Eqs. (\ref{eps_c2})
and (\ref{mu_c}). The chemical potential $\mu$ is
given in units of $k_{\rm B}T$.} 
\label{fig_pd}
\end{figure}

Finally, the binding phase diagram arising from Eq.~(\ref{f_mf2})
in the limit $\lp\gg 1$ is depicted in Fig.~\ref{fig_pd}. At the
lower limit of model validity, $\eps\rightarrow -1$, the spinodal
approaches a finite value, $\mu_{\rm sp}=\log(2/3)-5/2\simeq
-2.91$, whereas the binodal diverges. Indeed, for $\eps\rightarrow
-1$ the free energy (\ref{f_mf2}) tends to $-\infty$ for
$\ph=1$, regardless of the value of $\mu$, and the binodal is thus
obtained at $\mu\rightarrow -\infty$. In this respect, the limit
$\eps=-1$ for the bound molecules is similar to the limit of zero
temperature
--- the induced interaction is so strong that the molecules condense for
any value of the chemical potential. Note that in this special
limit, $\eps\rightarrow -1, \ph\rightarrow 1$, the effective
stiffness, $\lp(1+\eps\ph)$, becomes vanishingly small. This limit
cannot be accurately treated within the continuum form of the
semiflexible polymer Hamiltonian \cite{negative_stiffness}.

Equations (\ref{eps_c2})--(\ref{C_mf2}) and the phase diagrams in
Fig.~\ref{fig_pd} summarize the results obtained so far. They
indicate that in cases of semiflexible polymers, where binding of
small molecules significantly affects local chain features, the
binding should be a very sharp process. For finite $\lp$ the slope
of the binding isotherm is finite, \ie the binding is always
continuous, yet for $\lp\sim 10^2$ like in DNA, the behavior will
be practically indistinguishable from a discontinuous phase
transition.

It should be borne in mind that the sharp binding, obtained
despite the one-dimensionality of the model, relies on the long
range of the induced interaction. A direct short-range interaction
between bound molecules could not produce a similar effect. Hence,
such a short-range interaction (\eg a nearest-neighbor
interaction), which was omitted in Eq.~(\ref{Z1}) for the sake of
brevity, does not have an important effect on the binding in the
domain of interest, \ie $\lp\gg 1$ and $|\eps|\gtrsim 1$.

\section{Chains under Tension}
%-----------------------------
\label{sec_tension}

In addition, we consider binding to semiflexible chains which are
subject to external tension. This scenario is relevant to recent
single-molecule manipulation experiments \cite{Chatenay,Feingold}.
Since the tension suppresses chain fluctuations, it is expected to
have a significant effect on the fluctuation-induced mechanism
presented in the preceding sections.

In order to incorporate the external tension into our model, a
term is to be added to the chain Hamiltonian [cf. Eq.~(\ref{Z1})]
\cite{HT_tension},
\begin{eqnarray}
  Z &=& \tr_{\{\ph_n=0,1\}} \int\prod_{n=1}^N \rmd
  \vecu_n \exp(-\cH-\cH_{\rm t})
  \nonumber\\
  \cH_{\rm t} &=& -\vect\cdot\sum_{n=1}^N \vecu_n,
\label{Z1_tension}
\end{eqnarray}
where $\cH$ has been defined in Eq.~(\ref{Z1}), and $\vect$ is the
exerted tension (in units of $k_{\rm B}T$ divided by monomer
length).

As in Sec.~\ref{model}, we begin with the previously studied
problem of a bare semiflexible chain, yet it is now a chain under
tension \cite{HT_tension,MarkoSiggia}. The additional tension term
has not changed the Gaussian form of the polymer part of $Z$. It
can be calculated, therefore, in a similar way to that of
Sec.~\ref{model}, yielding
\begin{equation}
  Z_{\rm pt}^{1/N} = Z_{\rm p}^{1/N} \exp(t^2/4\lambda),
\end{equation}
where $Z_{\rm p}$ is the tensionless polymer partition function
given in Eq.~(\ref{Zp2}). The equation for the multiplier
$\lambda$ is, in this case,
\begin{equation}
  \half\left(\frac{3}{\lp\lambda}\right)^{1/2} +
  \frac{t^2}{4\lambda} = 1,
\label{lambda_sol_tension}
\end{equation}
which reduces to Eq.~(\ref{lambda_sol}) for $t=0$. The resulting
polymer free energy is
\begin{equation}
  f_{\rm pt} = \frac{3}{2}\log\lp +
  \left(\frac{3\lambda}{\lp}\right)^{1/2} -
  \frac{t^2}{4\lambda} - \lambda,
\label{fpt}
\end{equation}
where $\lambda=\lambda(\lp,t)$ is the solution to
Eq.~(\ref{lambda_sol_tension}).

For $\lp t\ll1$, the solution for $\lambda$ is
\[
  \lambda \simeq \frac{3}{4\lp} \left[ 1 + \frac{8}{9}(\lp t)^2 +
  \cO(\lp t)^4 \right],
\]
and the free energy becomes
\begin{equation}
  f_{\rm pt} \simeq f_{\rm p} - \frac{\lp}{3}t^2 + \cO(\lp^3 t^4),
  \ \ \ \ t \ll 1/\lp,
\label{fpt_weak}
\end{equation}
where $f_{\rm p}$ is the tensionless free energy given in
Eq.~(\ref{fp}). This is the elastic regime, where the energy is
quadratic (\ie the relative chain extension is linear) in tension
\cite{HT_tension,dG_book_tension}. Since we assume a large
persistence length, this regime corresponds to very weak tension,
$t\ll 1/\lp \ll 1$. In the opposite limit, $\lp t \gg 1$, the
solution to Eq.~(\ref{lambda_sol_tension}) becomes
\[
  \lambda \simeq \frac{t}{2} \left[ 1 + \half\left(\frac{3}{2\lp
  t}\right)^{1/2} + \cO(\lp t)^{-1} \right],
\]
and the corresponding free energy is
\begin{equation}
  f_{\rm pt} \simeq \frac{3}{2}\log\lp - t +
  \left(\frac{3t}{2\lp}\right)^{1/2} + \cO(\lp^{-1}t^0),
  \ \ \ \ t \gg 1/\lp.
\label{fpt_strong}
\end{equation}
In this regime the chain extension changes like the inverse square
root of tension \cite{HT_tension,Odijk}.

\begin{figure}
\centerline{ \epsfxsize=8.6cm \hbox{\epsffile{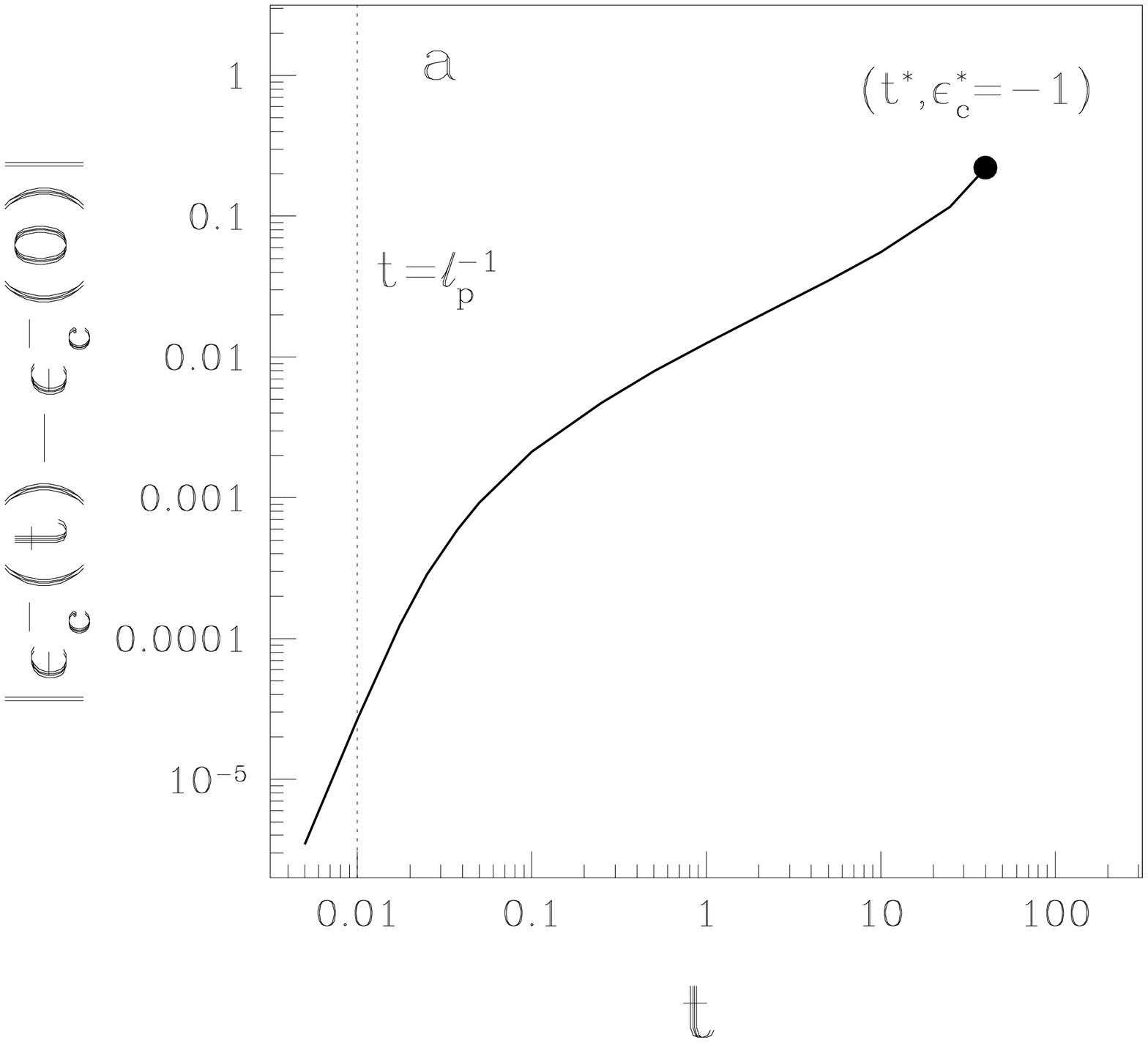}}
\epsfxsize=8.6cm \hbox{\epsffile{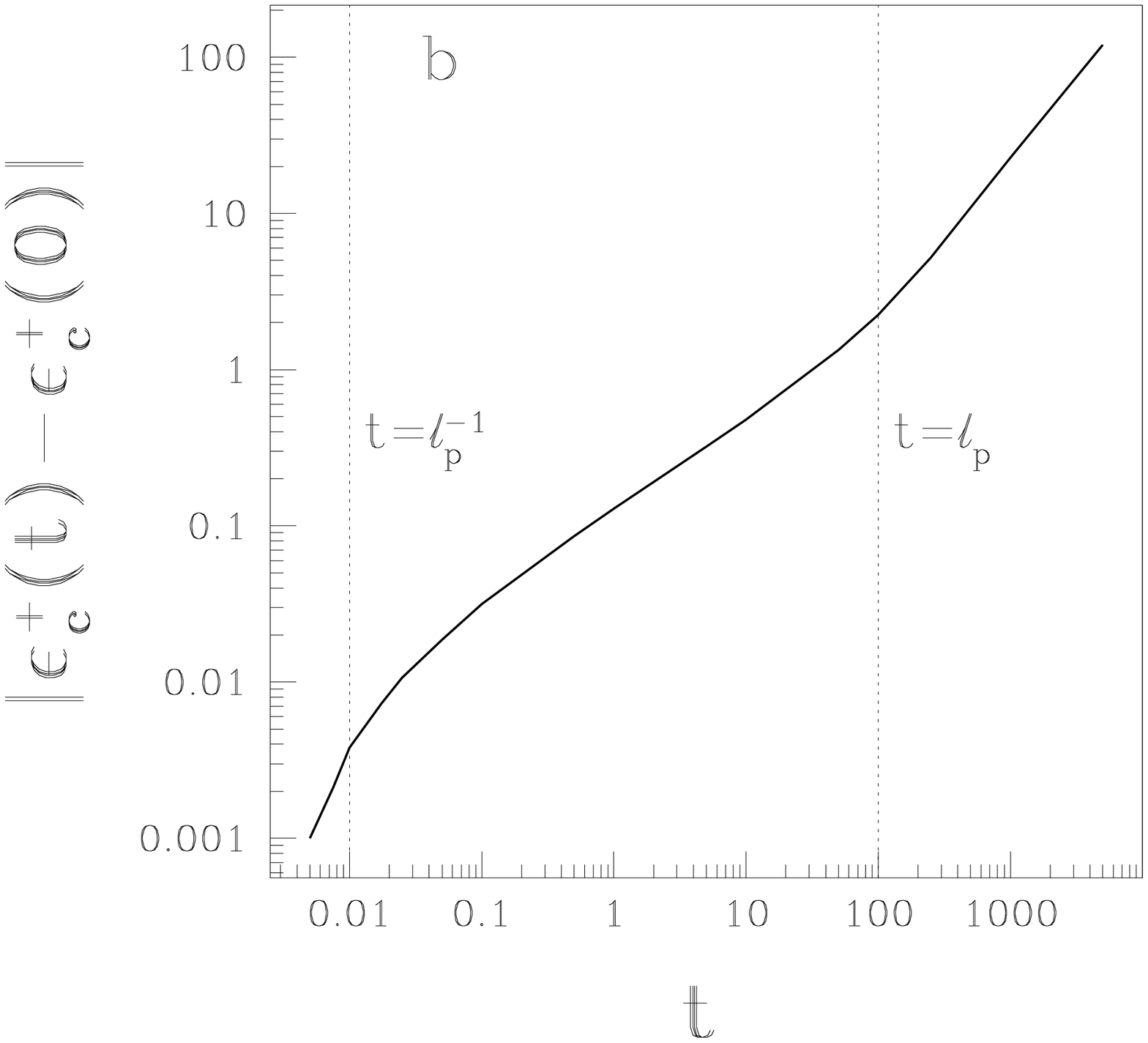}}}
%\centerline{
%\resizebox{0.5\textwidth}{!}{\includegraphics{fig4a.ps}}
%\resizebox{0.5\textwidth}{!}{\includegraphics{fig4b.ps}}}
\caption[]{Effect of tension on the critical behavior of binding.
(a) Critical coupling, $\eps_{\rm c}^-(t)<0$, as function of
tension. Two regimes are found: for $t\lesssim 1/\lp$,
$|\eps_{\rm c}^- (t)|$ increases like $t^4$; for
$t\gtrsim 1/\lp$, it increases like $t^{1/2}$. The critical line
terminates at the point $(t^*\simeq 0.410 \lp,\eps_{\rm c}^*=-1)$,
beyond which a sharp binding transition becomes unattainable. (b)
Critical coupling, $\eps_{\rm c}^+(t)>0$, as function of tension.
Apart from the two regimes of (a) there is a third one for
$t\gtrsim\lp$, where $\eps_{\rm c}^+ (t)$ increases linearly with
$t$. The value taken for $\lp$ in the numerical calculation is
100. The tension $t$ is
given in units of $k_{\rm B}T$ divided by monomer length.} 
\label{fig_ec_tension}
\end{figure}

Let us turn now to the effect of tension on the system of polymer
and bound molecules, Eq.~(\ref{Z1_tension}). As in
Sec.~\ref{strong}, we employ the mean-field approximation, valid
for $\lp\rightarrow\infty$. The resulting free energy is the same
as Eq.~(\ref{f_mf2}), but with $f_{\rm pt}$ instead of $f_{\rm
p}$,
\begin{equation}
  f = \left.f_{\rm pt}\right|_{\lp\rightarrow\lp(1+\eps\ph)}
  + \ph\log\ph + (1-\ph)\log(1-\ph) - \mu\ph.
\label{f_mf_tension}
\end{equation}
Due to the additional degree of freedom, namely tension, the
binding phase diagrams of Fig.~\ref{fig_pd} become
three-dimensional. In particular, the critical points $\eps_{\rm
c}^\pm$ become critical lines, $\eps_{\rm c}^\pm (t)$. (Note that
$\vect$ is an external field coupled to $\{\vecu_n\}$ rather than
$\{\ph_n\}$, and, hence, it does not destroy the critical
behavior.) The `condensation' of bound molecules in our model
results from attraction induced by polymer fluctuations. By
suppressing fluctuations, the tension should weaken the attraction
and shift the critical coupling to higher values, \ie increase the
positive critical point, $\eps_{\rm c}^+$, and decrease the
negative one, $\eps_{\rm c}^-$. Using Eqs.
(\ref{lambda_sol_tension}), (\ref{fpt}) and (\ref{f_mf_tension}),
the critical lines, $\eps_{\rm c}^\pm (t)$, can be calculated. The
results are shown in Fig.~\ref{fig_ec_tension}.

Before getting into the detailed effect of tension, we address the
question whether the critical behavior can survive {\em any}
strength of tension. In this respect there is an essential
difference between stiffness-strengthening binding ($\eps>0$) and
stiffness-weakening one ($\eps<0$). In the former case, since the
value of $\eps$ is unbound, there exists $\eps_{\rm c}^+ (t)$ for
any value of $t$, such that the binding is a sharp transition for
$\eps>\eps_{\rm c}^+ (t)$. In other words, the critical line
$\eps_{\rm c}^+ (t)$ exists for any $0\leq t<\infty$. Indeed,
substituting $\eps\rightarrow\infty$ in Eq.~(\ref{f_mf_tension})
while using Eq.~(\ref{fpt_strong}), we find that the free energy
always describes a sharp transition, regardless of the value of
$t$. On the other hand, in the latter case of stiffness-weakening
binding, there is a lower bound for $\eps$, $\eps>-1$, where the
validity of the entire approach breaks (see previous section).
Substituting $\eps=-1$ in Eqs. (\ref{f_mf_tension}) and
(\ref{fpt_strong}), we find that a critical point exists only for
$t<t^*$, where
\begin{equation}
  \frac{t^*}{\lp} = \frac{4}{9} \left( 33-7\sqrt{21} \right)
  \simeq 0.410.
\end{equation}
Thus, the critical line $\eps_{\rm c}^- (t)$ terminates at the
point $(t^*,\eps_{\rm c}^*=-1)$, beyond which a sharp binding
transition cannot be attained. This situation is similar to a case
where the critical temperature $T_{\rm c}$ coincides with $T=0$
(\eg in a one-dimensional Ising model), and the system is
disordered at all temperatures $T>0$.

Several regimes are found as function of tension. For very weak
tension, $t<1/\lp$, the leading-order term which couples
binding and tension is found from Eqs. (\ref{fpt_weak}) and
(\ref{f_mf_tension}) to scale like $\lp t^2\eps\ph$, \ie it is
only linear in $\ph$. Hence, to leading order in $\lp t$ there is
no effect on the critical point. Although the tension influences
chain fluctuations (\eg causing the chain to extend linearly with
$t$), it is too weak to affect the fluctuation-induced
interactions between bound molecules. The next-order term scales
like $\lp^3 t^4(1+\eps\ph)^3$, leading to a very small shift of
$\sim\lp^3 t^4$ in the critical point (see also
Fig.~\ref{fig_ec_tension}).

For $t>1/\lp$, the leading-order term in the free energy,
according to Eqs. (\ref{fpt_strong}) and (\ref{f_mf_tension}), is
$\sim(t/\lp)^{1/2}(1+\eps\ph)^{-1/2}$. Here two regimes should be
distinguished. For intermediate tension, $1/\lp<t<\lp$, the
critical line scales like $(t/\lp)^{1/2}$, reflecting a more
significant, yet still weak effect of tension. Although the chain
conformation is significantly stretched by tension in this regime,
the induced interaction between bound molecules is not strongly
affected. However, for $t>\lp$, the tension term in the free
energy [$\sim(t/\lp)^{1/2}(1+\eps\ph)^{-1/2}$] becomes dominant,
leading to a linear dependence of the critical point on tension,
$\eps_{\rm c}^+\sim t/\lp$.

The above analysis for the dependence of the critical coupling on
tension is summarized in the following expression:
\begin{equation}
  |\eps_{\rm c}^\pm (t) - \eps_{\rm c}^\pm (0)| ~\sim~ \left\{
  \begin{array}{ll}
  \lp^3 t^4 \ \ \ \ \ \ \ \ \ \  & t<1/\lp \\ & \\
  (\lp/t)^{1/2} & 1/\lp<t<\lp \\ & \\
  \lp/t & t>\lp, ~\mbox{relevant only to $\eps_{\rm c}^+$}.
  \end{array} \right.
\end{equation}
The various regimes are also clearly seen in
Fig.~\ref{fig_ec_tension}. Note that for the large values of $\lp$
considered in this theory the intermediate tension region,
$1/\lp<t<\lp$, is very wide.

\section{Discussion and Conclusions}
%-----------------------------------
\label{conclusions}

We have considered binding of small molecules to isolated
semiflexible polymer chains, where the persistence length $\lp$ is
much larger than the monomer size but still smaller than the total
chain length $N$. We have demonstrated that in such systems
polymer fluctuations induce attraction between bound molecules.
The long range of this interaction (of the same order as the
persistence length) can lead to strong effects on the binding
process. In particular, if bound molecules significantly affect
local features of the chain, \eg weaken or strengthen the
stiffness by a factor of about 5 ($\eps<\eps_{\rm c}^-$ or
$\eps>\eps_{\rm c}^+$), then the binding is predicted to be
extremely cooperative, occurring as a transition for a sharply
defined solute concentration. This is an unusual, yet practical
example for a one-dimensional system exhibiting a sharp transition
due to long-range interactions. The results of the model should
apply, in particular, to the association of DNA with smaller
molecules such as surfactants and compact proteins.

Subjecting the polymer to external tension has been studied as
well. By suppressing the fluctuation-induced interaction, the
applied tension may strongly affect the binding. The effect is
significant for sufficiently strong tension of order $t\sim\lp$.
[For DNA this implies $t\sim 10^2 k_{\rm B}T/(10 \mbox{\AA}) \sim
10^2$ pN.] In cases where binding weakens the chain stiffness,
such a high tension should make the sharp binding transition
disappear altogether (\ie regardless of the strength of coupling
or temperature). In cases where binding strengthens the chain
stiffness, a tension of $t\gtrsim\lp$ significantly shifts $\eps_{\rm
c}^+$ to higher values.
It is worth mentioning that tension-induced pairwise interaction 
between {\it specifically} bound proteins on a DNA chain was 
studied in a previous work \cite{Rudnick}.

The interaction of DNA with oppositely charged cationic
surfactants has been thoroughly studied by potentiometric
techniques \cite{Hayakawa,Shirahama} and fluorescence microscopy
\cite{Sergeyev1,Sergeyev5}. Isotherms measured by potentiometry
reveal a very cooperative, albeit continuous binding. Fluorescence
microscopy convincingly demonstrated, however, that the binding to
a {\em single} DNA molecule has a discrete nature resembling a
1st-order phase transition. It is usually accompanied by a
coil-to-globule collapse of the DNA chain (which lies outside the
scope of the current theory). The smoothness of potentiometric
isotherms was shown to arise from averaging over an {\em ensemble}
of DNA molecules, coexisting in bound and unbound
states \cite{Sergeyev1}. %\cite{Sergeyev2}--\cite{Sergeyev4}.
 Similar results were obtained for the association of DNA with
spermidine \cite{Khokhlov}. The microscopic origin of the observed
cooperativity (or even discontinuous transition) has not been
clarified. It is usually fitted to a phenomenological parameter
describing strong interaction between nearest-neighboring bound
molecules \cite{Ising_models}. On the other hand, it is reasonable
to expect that oppositely charged surfactants bound to DNA chains
significantly modify the chain stiffness (probably weakening it).
Thus, our model demonstrates that the strong cooperativity
observed in experiments can be well accounted for by weak, yet
long-range interactions induced by polymer fluctuations.

Recently, the kinetics of non-specific binding of RecA proteins to
DNA has been studied by single-molecule manipulation
\cite{Chatenay,Feingold}. RecA is a bacterial protein involved in
DNA recombination and known to cause significant changes in the
local structure of the double strand upon binding \cite{Stasiak}.
It was found to increase the DNA stiffness by a large factor,
estimated around 10 in one study \cite{Chatenay} and above 4 in
another \cite{Feingold}. This corresponds to a large, positive
$\eps$ in our model. A very cooperative nucleation-and-growth
kinetics was observed, as expected from the current model.
Moreover, in certain situations it was possible to achieve a
smaller increase of stiffness by binding of RecA. This led,
correspondingly, to a less cooperative process \cite{Feingold}.
Yet probably the most compelling evidence is that the binding
cooperativity was shown to be sensitive to external tension of
order 10--100 pN. It was consequently deduced that DNA
conformational fluctuations play a key role in RecA binding
\cite{Chatenay}, in accord with the model.

The current work is restricted to one-dimensional interactions
along the chain sequence, assuming that the polymer is locally
stiff and obeys the worm-like-chain description. Apart from
changing local properties of the polymer, an important feature not
treated by the model is that bound molecules may also modify {\em
volume} interactions between the monomers, thus affecting the
three-dimensional conformation of the polymer. For example,
binding of oppositely charged surfactants to a DNA molecule
locally neutralizes the DNA charge. This should lead, indeed, to a
modified stiffness, but also to a reduced 2nd virial coefficient,
which may drive a coil-to-globule collapse \cite{Sergeyev1}. The
collapse can be also driven by fluctuations in the concentration
of ions adjacent to the chain, as has been demonstrated by recent
theoretical studies \cite{Parsegian,Kardar}.

In order to check the theory presented in this work more
experiments are required, focusing, in particular, on the effect
of persistence length and tension on binding. The fluorescence
microscopy techniques, which have been successfully used for
DNA--surfactant association, may be applied to chains under
tension or flow, thus examining the role of fluctuations. It
would be interesting to study a system consisting of a
semiflexible polymer and bound molecules in computer simulations,
and thereby check the applicability of our mean-field
approximation. An important extension of the model, as mentioned
above, would be to introduce volume interactions and obtain
binding-induced collapse as observed in experiments.

\acknowledgments

We greatly benefited from discussions and correspondence
with R. Bar-Ziv, M. Feingold, A. Libchaber, R. Netz, A. Parsegian, 
R. Podgornik, M. Schwartz and V. Sergeyev. 
Partial support from the Israel Science Foundation
founded by the Israel Academy of Sciences and Humanities ---
Centers of Excellence Program, and the Israel--US Binational
Science Foundation (BSF) under grant No. 98-00429, 
is gratefully acknowledged. HD would
like to thank the Clore Foundation for financial support.

\appendix

\section*{Numerical Details}
%-------------------------------------

\begin{figure}
\centerline{ \epsfxsize=8.6cm \hbox{\epsffile{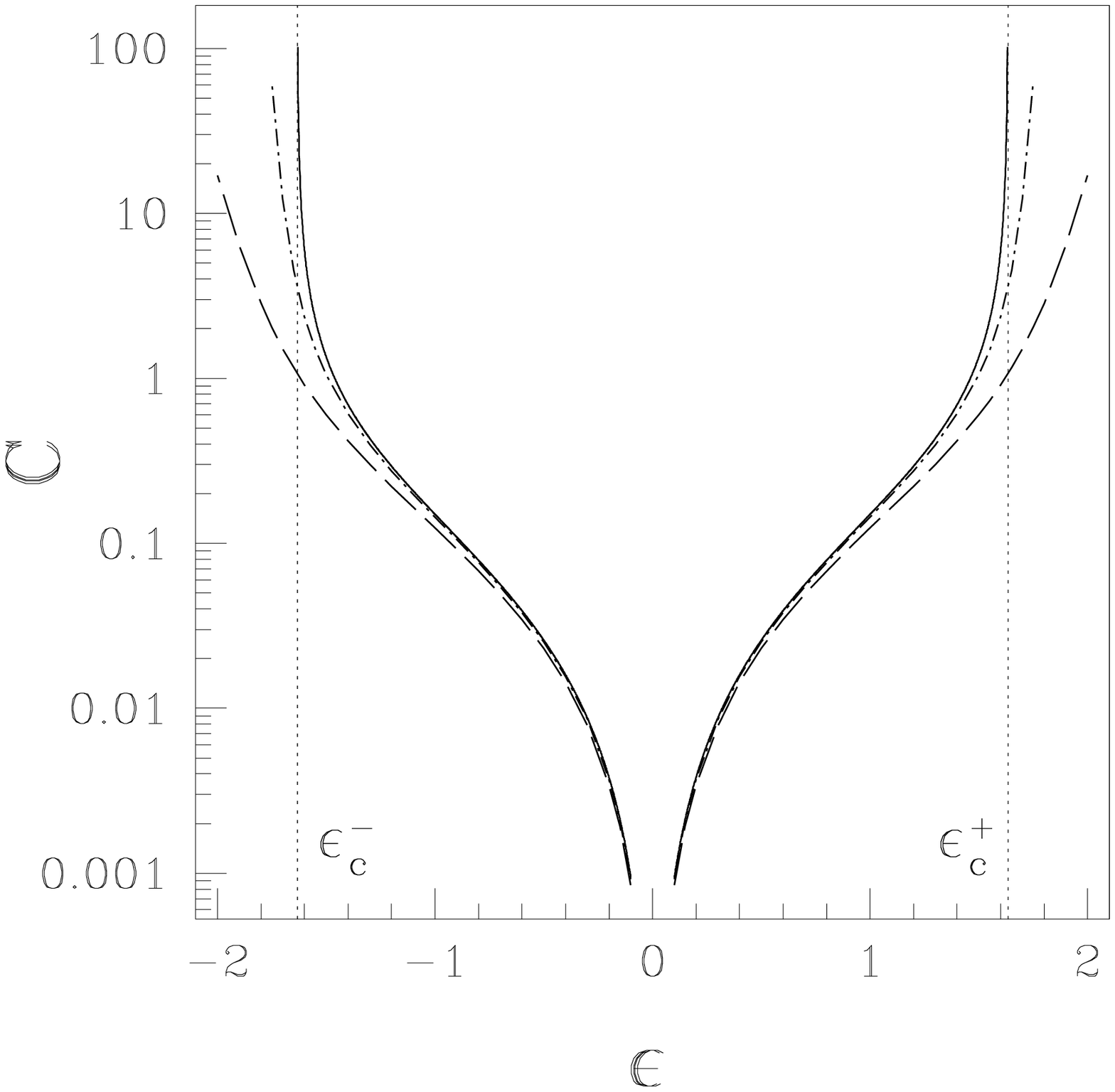}}}
%\centerline{ \resizebox{0.6\textwidth}{!}
%{\includegraphics{fig1.ps}}}
\caption[]{Binding cooperativity as function of $\eps$ according
to the Kac-Baker model, plotted on a semi-logarithmic scale. The
dashed and dash-dotted curves are results of numerical
calculations for $\lp=10$ and 50, respectively. The solid curves
show analytic results for $\lp\rightarrow\infty$ as obtained by a
mean-field calculation [Eq.~(\ref{C_mf1})]. The critical points
are at $\eps_{\rm c}^\pm=\pm\sqrt{8/3}$ (dotted lines).}
\label{fig_Kac}
\end{figure}

The aim of the numerical scheme is to calculate the results of the
Kac-Baker model for finite $\lp$, which are presented in
Fig.~\ref{fig_Kac}. Using Kac's solution \cite{Lieb,Kac}, the
partition function of bound solute molecules,
Eqs.~(\ref{Z4})-(\ref{vmn}), is expressed in the limit
$N\rightarrow\infty$ as
\begin{equation}
  Z_{\rm s} = \mbox{const}\times e_0^N,
\end{equation}
where $e_0$ is the largest
eigenvalue of the following `transfer kernel':
\begin{equation}
  K(x,y) = [(1+\rme^{\mu-3\eps/2+\sqrt{J}x})
  (1+\rme^{\mu-3\eps/2+\sqrt{J}y})]^{1/2} \exp\left[
  \frac{y^2-x^2}{4} - \frac{(y-\rme^{-2/\lp}x)^2}
  {2(1-\rme^{-4/\lp})} \right],
\label{kernel}
\end{equation}
where $J\equiv 3\eps^2/2\lp$, and $x,y\in(-\infty,\infty)$
are real variables.

We define a vector,
$\{x_i\}=\{(2i-M)d\}_{i=0\ldots M}$, where $M$ is an even
integer and $d$ a real number, and use it to discretize the
kernel $K(x,y)$ into a transfer matrix,
\begin{equation}
  K_{ij}\equiv K(x_i,x_j).
\end{equation}
In addition, we define the diagonal matrix
\begin{equation}
  A_{ij}\equiv x_i\delta_{ij}.
\end{equation}
Given $\lp$, $\eps$ and $\mu$, the transfer matrix $K_{ij}$ is
diagonalized and its largest eigenvalue, $e_0$, is found.

The binding degree, $\ph$, can be calculated in two ways.
The first is by calculating the variation of $\log e_0$ with
respect to $\mu$,
\begin{equation}
  \ph = \partial\log e_0/\partial\mu.
\end{equation}
The second way is by using the equation
\begin{equation}
  \ph = \tilde{A}_{00}/(B\sqrt{J}),
\end{equation}
where $B\equiv\coth(1/\lp)$, and $\tilde{A}$
is the matrix $A$ transformed to the basis
where $K$ is diagonal \cite{critical}.

The cooperativity parameter, $C$, as defined in
Eq.~(\ref{cooperativity1}), is found by calculating the variation
of $\ph$ with respect to $\mu$ around the point $\ph=1/2$. The
value $\mu=\mu_{1/2}$ which gives $\ph=1/2$ is analytically found
by transforming the lattice-gas partition function,
Eqs.~(\ref{Z4})-(\ref{vmn}), into an Ising one ($\ph_n\rightarrow
s_n=2\ph_n-1$), and requiring that the `magnetic field'
coefficient should vanish. The result is
\begin{equation}
  \mu_{1/2}=3\eps/2 - JB/2.
\end{equation}

For each calculation (\ie for each set of $\lp$, $\eps$ and
$\mu$) the discretization parameters, $M$ and $d$,
were tuned until the result became insensitive to further
refinement to six significant figures.
In addition, the two methods for calculating $\ph$ were
used and verified to yield identical results to six figures.
All algebraic manipulations were performed using
{\it Mathematica}.

%------------------------------------------------------

%------------------------------------------------------

\pagebreak

\pagebreak

\pagebreak

%------------------------------------------------------

\begin{references}
\bibitem{Rudnick2}
That is also the reason why stiffness-modifying binding,
considered by a previous work (Ref.~\cite{Rudnick}),
did not produce any cooperativity effect without external
tension.
\bibitem{proof}
Note that this statement, though plausible, is not rigorously
proved. For values of $\eps$ which are not very small, the
cumulant expansion of Eq.~(\ref{Z2}) would produce higher-order,
many-body terms, whereas the proof by Lebowitz and Penrose, Ref.
\cite{Lebowitz}, applies to pair potentials only. On the other
hand, all those higher-order terms have ranges proportional to
$\lp$ (it is the only length scale in the problem when $N$ is
infinite) and, hence, should satisfy the mean-field assumption for
$\lp\rightarrow\infty$.
\end{references}
\end{document}